\documentclass[12pt]{article}  % DEFINIÇÃO QUE RESOLVEU  O  PROBLEMA COM A  FIGURA

\usepackage{graphicx,color,epsf} % PACOTE PARA INSERIR FIGURAS

\usepackage{amsmath,latexsym}

\newcommand{\gsim}{\raisebox{-0.13cm}{~\shortstack{$>$ \\[-0.07cm]
      $\sim$}}~}
\newcommand{\lsim}{\raisebox{-0.13cm}{~\shortstack{$<$ \\[-0.07cm]
      $\sim$}}~}
\begin{document}

%\begin{flushright}
%QMW/95-5
%\\hep-th/9502016
%\end{flushright}
%\vspace{0.5cm}
%%%%%%%%%%%%%%%%%%%%%%%%%%%%%%%%%%%%%%%%%%%%%Carol new address%%%%%%%%%%
%%%%%%%%%%%%%%%%%%%%%%%%%%%%%%%%%%%%%%
\begin{center}
	{\LARGE \bf
		Thermodynamics of a charged relativistic ideal Boltzmann gas
	}
	\\
	\vspace{1cm}
	{\large  C. P. Felix$^{a}$ and E. S. Moreira Jr.}$^{b,}$\footnote{E-mail:
		{\tt felix.carol@nsrrc.org.tw} \& {\tt moreira@unifei.edu.br (corresponding author)}}
	\\
	\vspace{0.3cm}
	{$^{a}$\em  Department of Beam Dynamics Group,}
	{\em National Synchrotron Radiation Research Center,}   \\
	{\em East District, Hsinchu 30076, Taiwan, ROC}  \\
	\vspace{0.1cm}
	{$^{b}$\em Instituto de Matem\'{a}tica e Computa\c{c}\~{a}o,}
	{\em Universidade Federal de Itajub\'{a},}   \\
	{\em Itajub\'a, Minas Gerais 37500-903, Brazil}
%%%%%%%%%%%%%%%%%%%%%%%%%%%%%%%%%%%%%%%%%%%%%%%%%%%%%%%%%%%%%%%%%%%%%%%%%%%%%%%%%%%

%\begin{center}
%{\LARGE \bf 
%Thermodynamics of a charged relativistic ideal Boltzmann gas
%}
%\\ 
%\vspace{1cm}
%{\large  C. P. Felix$^{a}$ and E. S. Moreira Jr.}$^{b,}$\footnote{E-mail:
%{\tt carolfba@gmail.com} \& {\tt moreira@unifei.edu.br }}
%\\
%\vspace{0.3cm}
%{$^{a}$\em Carol's Institute,}
%{\em Carol's University,}   \\
%{\em Carol's City, Taiwan}  \\
%\vspace{0.1cm}
%{$^{b}$\em Instituto de Matem\'{a}tica e Computa\c{c}\~{a}o,}
%{\em Universidade Federal de Itajub\'{a},}   \\
%{\em Itajub\'a, Minas Gerais 37500-903, Brazil} 
%\vspace{1cm} 
%{\large E. S. Moreira Jr.}
%\footnote{E-mail: moreira@unifei.edu.br}   
%\\ 
%\vspace{0.3cm} 
%{%\em Instituto de Matem\'{a}tica e Computa\c{c}\~{a}o,}  
%{\em Universidade Federal de Itajub\'{a},}   \\
%{\em Itajub\'a, Minas Gerais 37500-903, Brazil}

\vspace{0.3cm}
%{\large April, 2019}
{\large November, 2022}
\end{center}
\vspace{0.5cm}

%\end{document}

%\begin{abstract}

\begin{abstract}
	
This paper presents a toy model of a charged relativistic classical gas in flat spacetime of an arbitrary number of dimensions equipped with some
``pair production'' mechanism. Working with the microcanonical ensemble, the
charge is taken to be conserved in contrast with the total number of ``particles'' plus ``antiparticles'' which varies with temperature $T$. Thermodynamics of the classical gas
is detailed studied in the nonrelativistic ($Mc^{2}\gg kT$) as well as in the ultrarelativistic ($Mc^{2}\ll kT$) regimes. 
It is shown that, although we are dealing with classical distributions, at the ultrarelativistic regime the behavior of the gas is Planckian. We also compare the thermodynamics of the toy model with that of quantum gases.

\end{abstract}
\vspace{0.1cm}
\hspace{0.9cm}

%\vspace{0.1cm}
%\hspace{0.3cm}
%PACS number(s): 11.10.Wx, 04.62.+v, 05.70.-a

\section{Introduction}
In 1928, Paul Dirac solved the negative energy problem in the Klein-Gordon equation by assuming that there was an object  resembling a particle but with an opposite electric charge \cite{dir28}. His theory would be experimentally confirmed  in 1932 by Carl Anderson with the discovery of the electron's antiparticle, the positron \cite{and33}.
Since then, many experimental and theoretical discoveries have shown that at extreme energies (e.g., either inside modern accelerators or in the early universe) pair production must be taken into account.

When one tries to extend classical statistical mechanics of an ideal gas with $N$ particles of mass $M$, in a cavity of volume $V$,
to the realm of relativity the prescription seems to be simply to consider that the particles have relativistic energies. Such a prescription (see, e.g., ref. \cite{and67})
results in  
%their model suffers from oversimplification.
the familiar equation of state:
\footnote{Fundamental constants have the usual meaning.}
\begin{equation}
PV=NkT,	
\label{boyle}	
\end{equation}
and either in the nonrelativistic internal energy:
\begin{equation}
U=N Mc^{2}+\frac{3}{2}NkT,\hspace{1.0cm}
Mc^{2}\gg kT,
\label{nru}
\end{equation}
or in the ultrarelatistic one:
\begin{equation}
	U=3NkT,\hspace{2.8cm} Mc^{2}\ll kT,
	\label{ru}
\end{equation}
with
both expressions for $U$ holding in leading order. In deriving eqs. (\ref{boyle}) to (\ref{ru}) it is assumed that $N$ is conserved, which is an oversimplification as far as eq. (\ref{ru}) is concerned.
And the reason is simple. For instance, if particles are electrons, as temperature grows eventually positrons will appear resulting that $N$ (the number of electrons plus the number of positrons) increases while the net charge $Q$ (the number of electrons minus the number of positrons) remains the same according to what quantum field theory at finite temperature prescribes \cite{kap06}. 

It is worth remarking that even 
quantum statistical mechanics with $N$ conserved would lead to eq. (\ref{ru}) for high enough $T$ \cite{car80}. The issue was resolved in 1981 by Haber and Weldon \cite{wel81} who considered an ideal Bose gas with conserved net charge $Q$ (the number of particles minus the number of antiparticles).
%, instead of $N$, 
Analogous considerations can be applied to an ideal Fermi gas
\cite{kap06,cai10}, resulting that for both quantum gases eq. (\ref{ru}) is replaced by
the following Planckian internal e\-ner\-gy,
\footnote{It is assumed that $T$ is high enough, i.e., much bigger than the Fermi temperature (for fermions) or bigger than the critical temperature (for bosons).}
\begin{equation}
	U=\sigma VT^{4},\hspace{2.8cm} Mc^{2}\ll kT,
	\label{ur}
\end{equation}
where the value of the ``Stefan-Boltzmann'' constant $\sigma$ 
depends on the nature of the gas
\footnote{$\sigma=\pi^{2}k^{4}/15(\hbar c)^{3}$
or $\sigma=7\pi^{2}k^{4}/60(\hbar c)^{3}$ for bosons and fermions, respectively.}.
It should be added that, corresponding to the ultrarelativistic regime in eq. (\ref{ur}),
now $P=\sigma T^{4}/3$.

After going through the literature, one can see that a
toy model which has been overlooked is that of a charged relativistic  ideal classical gas equipped with some mechanism of ``pair production'', which could be of any nature. It would be interesting to compare the associated  thermodynamics with that of charged quantum gases, identifying 
%its 
limitations of the toy model. And this is the aim of the present paper whose organization is as follows. 
For the sake of generalization, we will work in  $D$-dimensional Minkowski spacetime. 

The next section is devoted to deriving equilibrium particle and antiparticle distributions by means of the microcanonical ensemble \cite{hua87}.
Two Lagrange multipliers arise corresponding to the charge $Q$ and energy $E$.
In Section \ref{thermodynamics},
 chemical potential $\mu$, temperature $T$ and pressure $P$ are obtained from the equilibrium entropy as a function of $U\equiv E$, $V$ and $Q$, where $V$ is the $(D-1)$-dimensional volume of the cavity containing an ideal gas with particles and antiparticles  of mass $M$.
%A plot illustrates how the relative densities of particles and antiparticles vary %with $T$ while another shows $\mu$ vs. $T$.
Thermodynamic quantities are expressed in terms of modified Bessel functions of 
the second kind, $K_{\nu}(Mc^{2}/kT)$, with $\nu$ determined by $D$. 
%For convenience a short appendix is included containing few properties of $K_{\nu}(z)$. 
In Sections \ref{nrregime}
and \ref{urregime}, respectively, thorough investigations of thermodynamics at the nonrelativistic ($Mc^{2}\gg kT$) and ultrarelativistic ($Mc^{2}\ll kT$) regimes 
are performed
%Such analysis is implemented 
by using  the leading order behaviors of $K_{\nu}(z)$ as $z$ is
big and small. In Section \ref{discussion}, we close with a comparison between the thermodynamics of our toy model with that of realistic quantum gases where
rather curious features are spotted.
For convenience, a short appendix is included containing a few properties of $K_{\nu}(z)$.

\section{Most probable distributions}
\label{distributions}
We begin with a
relativistic classical ideal gas with charge
\begin{equation}
\label{charge}
	Q=N_+ -N_-,  
\end{equation}
where $N_+$ is the number of particles and $N_-$ is the number of antiparticles. If 
$n_{(+)i}$ and $n_{(-)i}$
are occupation numbers of the 
{\it i}\,th cell of the phase space with $K$ cells, then
\begin{equation}
	\label{particle_anti}
	N_+=\sum_{i=1}^{K}n_{(+)i},  \qquad   N_-=\sum_{i=1}^{K}n_{(-)i}.
\end{equation}
To each set of occupation numbers
$\{(n_{(+)i},n_{(-)i})\}$
corresponds to a certain number of states
$\Omega$, and the total number of states is given by 
%\left(\{(n_{(+)i},n_{(-)i})\}\right)$
\begin{equation}
	\label{space_volume}
	\Gamma = 
	\sum_{\{(n_{(+)i},n_{(-)i})\}}
	\Omega\left(\{(n_{(+)i},n_{(-)i})\}\right).  
\end{equation}
%contributes overwhelmingly
Considering the number of ways one can arrange distinguishable objects into $K$ cells, it follows that
\cite{hua87}
\begin{equation}
	\label{omega_r}
	\Omega
	%(\{n_{(+)i}!n_{(-)i}!\})
	\propto \prod_{i=1}^{K}\frac{1}{n_{(+)i}!n_{(-)i}!},
\end{equation}
where the ``correct Boltzmann counting'' has been used
\footnote{That is, original $\Omega$ has been divided by $N_+!\, N_-!$.}.

The distributions 
$\overline{n}_{(+)i}$ and $\overline{n}_{(-)i}$ 
for which $\ln\Omega$
is maximum will now be determined.
Two constraints will be taken into account
along with their Lagrangian multipliers $\alpha$ and $\beta$. One is the net charge in eq. (\ref{charge}) [see also eq. 
(\ref {particle_anti})], and the other is the gas
total energy:
\begin{equation}
	%\label{energia_fixa_r}
	E=\sum_{i=1}^{K}[n_{(+)i}+n_{(-)i}]\epsilon_{i},   
%\end{equation}
\hspace{1.0cm}
%\begin{equation}
	\label{epsilon}
	\epsilon_i=c\sqrt{M^2c^2+p_i^2}.
\end{equation}
Thus, by noting eqs. (\ref {charge}),
(\ref {particle_anti}), (\ref{omega_r}) and (\ref{epsilon}),
the solution of
\begin{equation}
	\delta\, \left[\textrm{ln}\, \Omega +\alpha\sum^{K}_{i=1}\left(n_{(+)i}-n_{(-)i}\right)-\beta\sum^{K}_{i=1}\left(n_{(+)i}+n_{(-)i}\right)\epsilon_{i}\right]
	\Bigg|_{\left(n_{(+)i},n_{(-)i}\right)=(\overline{n}_{(+)i},\overline{n}_{(-)i})}=0
\nonumber	
	%\label{variacao}
\end{equation}
%variation
%of $\ln\Omega$ leads to
is given by:
\begin{equation}
\overline{n}_{(+)i}=e^{\alpha}e^{-\beta\epsilon_{i}},
%\label{n_+}
%\end{equation}
%and
\hspace{1.0cm}
%\begin{equation}
\overline{n}_{(-)i}=e^{-\alpha}e^{-\beta\epsilon_{i}},
\label{n}
\end{equation}
where Stirling's approximation,
%\footnote{$\textrm{ln}\, n!=n\, \textrm{ln}\, n-n+\cdots$.}
\begin{equation}
\label{stirling}
\textrm{ln}\, n!=n\, \textrm{ln}\, n-n+\cdots,	
\end{equation}
 has been considered as usual.
%\footnote{$\textrm{ln}\, n!=n\, \textrm{ln}\, n-n+\cdots$.}.
%has been used as usual.

Before associating $\alpha$ and $\beta$
in eq. (\ref{n}) with thermodynamic parameters,
let us covert summations over $i$ 
into integrations. This can be achieved by
taking 
%assuming that 
an arbitrary small volume in phase space, denoted by
$h^{D-1}$, where the constant $h$ has dimensions of 
distance $\times$ momentum. For example, corresponding
to the most probable distributions in eq. (\ref{n}),
eq. (\ref{particle_anti}) yields
\begin{equation}
	\label{particle_antiparticle}
N_+=e^{\alpha}\frac{V}{\Lambda^{D-1}},  \qquad   N_-=e^{-\alpha}\frac{V}{\Lambda^{D-1}},
\end{equation}
with
\begin{equation}
	\Lambda^{-(D-1)}=\frac{1}{h^{D-1}}\int e^{-\beta c\sqrt{M^{2}c^{2}+p^{2}}}d^{D-1}p.
	\label{tlength1}
\end{equation}
Integrations in eq. (\ref{tlength1}) run from
$-\infty$ to $\infty$ over  $D-1$ components of the Euclidean vector ${\bf p}$ whose squared norm is $p^{2}$. We will call $\Lambda$ ``thermal wavelength'' though it is clearly an abuse of language once we are dealing with a classical gas.
Shortly a closed formula for $\Lambda$ will be determined.
%in eq. (\ref{tlength1}).
Noting eq. (\ref{particle_antiparticle}) one sees that $Q$ in eq. (\ref{charge}) 
and 
\begin{equation}
\label{totaln}	
N=N_{+} +N_{-}
\end{equation}
can also be written in terms of $\Lambda$, i.e., 
\begin{equation}
	Q=2\sinh(\alpha)\frac{V}{\Lambda^{D-1}},
	\hspace{1.0cm}
	N=2\cosh(\alpha)\frac{V}{\Lambda^{D-1}},
	\label{qn}
\end{equation}
which gives rise a neat relation, namely:
\begin{equation}
	N=\sqrt{Q^{2}+\frac{4V^{2}}{\Lambda^{2(D-1)}}}.
	\label{N}
\end{equation}

The total energy in eq. (\ref{epsilon}) has the following expression in terms of 
$\Lambda$, more precisely in terms of its derivative:
\begin{equation}
E=-2\cosh(\alpha)V\frac{\partial}{\partial \beta} \Lambda^{-(D-1)},
\label{ienergy}	
\end{equation}
as can be quickly checked by looking at eqs. (\ref{n}) and (\ref{tlength1}).
%We will shortly determine a close formula for $\Lambda$ in eq. (\ref{tlength1}).

Considering eqs. (\ref{particle_antiparticle}), (\ref{totaln}) and the densities
\begin{equation}
\label{densities}	
n_{+}=\frac{N_{+}}{V}, \hspace{1.0cm}n_{-}=\frac{N_{-}}{V}, \hspace{1.0cm}n=\frac{N}{V},
\end{equation}
%$n_{+}=N_{+}/V$, $n_{-}=N_{-}/V$ and $n=N/V$,
it follows relative densities:
\begin{equation}
	\frac{n_{+}}{n}=\frac{1}{e^{-2\alpha}+1},
	%\label{rdensity}
	\hspace{1.0cm}
\frac{n_{-}}{n}=\frac{1}{e^{2\alpha}+1},
\label{rdensity}	
\end{equation}
to which we will return later in the text.
Notice that $n_{+}/n+n_{-}/n=1$ as should be.

We turn now to determining a closed form for the
``thermal wavelength'' in eq. (\ref{tlength1})
in terms of modified Bessel functions of the second kind 
(see Appendix \ref{appendix}).
Spherical polar coordinates (see, e.g., ref. \cite{hua87}),
allow us to write,
\begin{equation}
	\Lambda^{-(D-1)}=\frac{2\pi^{(D-1)/2}}
	{h^{D-1}\Gamma\left(\frac{D-1}{2}\right)}
	\int_{0}^{\infty} p^{D-2}e^{-\beta c\sqrt{M^{2}c^{2}+p^{2}}}dp.
	\label{tlength2}
\end{equation}
Setting $p=Mc\sinh(\omega)$, eq. (\ref{tlength2}) yields
\begin{equation}
	\Lambda^{-(D-1)}=\frac{2\pi^{(D-1)/2}}
	{h^{D-1}\Gamma\left(\frac{D-1}{2}\right)}(Mc)^{D-1}
	\int_{0}^{\infty} \cosh(\omega)\sinh^{D-2}(\omega)
	e^{-\beta Mc^{2}\cosh(\omega)}d\omega.
	%\label{tlength3}
	\nonumber
\end{equation}
Thus, 
%the integral representation of the modified Bessel function of the second kind in
eq. (\ref{Macdonald2}) leads to:
\begin{equation}
	\Lambda^{-(D-1)}=\frac{{2}^{D/2}\pi^{(D-2)/2}}
	{h^{D-1}}
	\beta^{(2-D)/2}
	M^{D/2}cK_{\frac{D}{2}}(\beta Mc^{2}).
	\label{tlength4}
\end{equation}

By looking at the expression for $N$ in eq. (\ref{qn}), eq. (\ref{tlength4}) can be readily used in eq. (\ref{ienergy}), resulting in
\begin{equation}
	E=N\beta^{-1}\left[D-1+\beta Mc^{2}\frac{K_{\frac{D}{2}-1}(\beta Mc^{2})}{K_{\frac{D}{2}}(\beta Mc^{2})}\right],
	\label{ienergy2}
\end{equation}
where the second identity in eq. (\ref{p1}) has been considered.

\section{Thermodynamics}
\label{thermodynamics}
Entropy is defined by $S=k\ln \Gamma$, with $\Gamma$ given in eq. (\ref{space_volume}). Since $\overline{n}_{(+)i}$ and $\overline{n}_{(-)i}$ 
contribute overwhelmingly, it follows that
\begin{equation}
\label{entropy}
S=k\ln\Omega\left(\{(\overline{n}_{(+)i},\overline{n}_{(-)i})\}\right).
\end{equation}
Now, neglecting an unimportant constant, by using eqs. (\ref{omega_r}) to (\ref{stirling}) and also eqs. (\ref{charge}), (\ref{particle_anti}) and (\ref{totaln}), we can show that $S$ in 
eq. (\ref{entropy}) can be recast as
\begin{equation}
\label{entropy2}
S=k\beta E-k\alpha Q+kN.	
\end{equation}
The Lagrangian multipliers $\alpha$ and $\beta$ can be associated with thermodynamic
quantities by taking into account that they are functions of $E$, $V$ and $Q$. Thus,
recalling  that $\Lambda=\Lambda(\beta)$, and noticing  eqs. (\ref{qn}), (\ref{ienergy}) and (\ref{entropy2}), the usual 
definitions
\begin{equation}
\frac{1}{T}=\left(\frac{\partial S}{\partial E}\right)_{V,Q},	
\hspace{1.0cm}
\mu=-T\left(\frac{\partial S}{\partial Q}\right)_{E,V}
%\label{definition1}
\nonumber
\end{equation}
lead to
\begin{equation}
\label{lmultipliers}
\beta=\frac{1}{kT},\hspace{1.0cm}
\alpha=\frac{\mu}{kT}.
\end{equation}
Likewise,
\begin{equation}
\nonumber
P=T\left(\frac{\partial S}{\partial V}\right)_{E,Q}
\end{equation}
yields eq. (\ref{boyle}); but now with the crucial difference that $N$ is not conserved
[cf. eq. (\ref{N})]. Putting these findings together, eq. (\ref{entropy2}) becomes
\begin{equation}
\label{entropy3}
U=TS-PV+\mu Q,
\end{equation}
where $E$ has been identified as the internal energy $U$, as usual.

An expression for the chemical potential in eq. (\ref{lmultipliers})
can be obtained by noting eq. (\ref{qn}). One sees that
\begin{equation}
	\mu=kT\, \textrm{arcsinh}\left(\Lambda^{D-1}\,\frac{Q}{2V}\right).
	\label{cpotential}
\end{equation}
It is worth remarking that $\mu$ can be recast as
\footnote{$\textrm{arcsinh}(x)=\ln\left(x+\sqrt{x^{2}+1}\right)$.}
\begin{eqnarray}
	\mu&=&\hspace{0.3cm} kT\, \textrm{ln}\left(\Lambda^{D-1}\,
	n_{+}\right),
	\label{cpotential2}
	\\
	&=&-kT\, \textrm{ln}\left(\Lambda^{D-1}\,n_{-}\right),
	\label{cpotential3}
\end{eqnarray}
where eqs. (\ref{charge}), (\ref{totaln}),  (\ref{N}) and (\ref{densities}) have been used.

It follows immediately from eqs. (\ref{lmultipliers})  and (\ref{cpotential}) that $\mu$ vanishes if and only if $\alpha=0$ and the gas is neutral (i.e., $Q=0$). Then the relative densities in eq. (\ref{rdensity}) are both equal to $1/2$. In other words when $\mu$ vanishes the numbers of particles and antiparticles are the same for all temperatures:
\begin{equation}
\label{neutralgas}	
n_{+}=n_{-}=\frac{n}{2}=\frac{1}{\Lambda^{D-1}},\hspace{1.5cm}\mu=0.
\end{equation}
In deriving eq. (\ref{neutralgas}) we have noted eqs. (\ref{N}) and (\ref{densities}).
%Considering the identity
%$\textrm{arcsinh}(x)=\ln\left(x+\sqrt{x^{2}+1}\right)$, 
Now we go into a more detailed study of thermodynamics by looking at asymptotic regimes of large and small masses. Unless stated otherwise,
only contributions due to the leading order terms in eq. (\ref{abehavior}) will appear in the formulae below.
%Before going into a more detailed study of thermodynamics, it is perhaps pedagogical
%to show plots giving a crude idea 
%illustrating  
%how some quantities vary with $T$ \cite{mat17}.
%Figs. 1 and 2 illustrate  thermal behaviors of $\mu$ and of the relative densities
%in eq. (\ref{rdensity}), respectively [note eq. (\ref{lmultipliers})].

%This last term is an abuse of language since

\subsection{Nonrelativistic regime: $Mc^{2}\gg kT$}
\label{nrregime}
By considering eqs. (\ref{tlength4}), (\ref{lmultipliers}) and the second expression in eq. (\ref{abehavior}),
one ends up with
\begin{equation}
	\Lambda^{D-1}=\lambda^{D-1}e^{\frac{Mc^{2}}{kT}},
	\hspace{1.0cm}
\lambda=\sqrt{\frac{h^{2}}{2\pi MkT}}.	
	\label{tlength5}
\end{equation}
Correspondingly, eqs. (\ref{cpotential2})  and (\ref{cpotential3}) yield
\begin{eqnarray}
	\mu&=&\hspace{0.3cm} Mc^{2}+kT\, \textrm{ln}\left(\lambda^{D-1}\,
	n_{+}\right),
	\label{cpotential4}
	\\
	&=&-Mc^{2}-kT\, \textrm{ln}\left(\lambda^{D-1}\,n_{-}\right),
	\label{cpotential5}
\end{eqnarray}
respectively. Either eqs. (\ref{cpotential4}) and (\ref{cpotential5})
or eq. (\ref{tlength5}) quickly show that for a neutral gas (i.e., $\mu=0$)
the densities in eq. (\ref{neutralgas}) and $P$ in eq. (\ref{boyle}) are exponential small.

Let us look more carefully at the densities when $Q\neq 0$. Then, eq. (\ref{N})
leads to 
\begin{equation}
	n=\frac{|Q|}{V}\sqrt{1+\frac{4V^{2}}{Q^{2}\Lambda^{2(D-1)}}}.
	\label{N2}
\end{equation}
%which holds for all regimes. 
For the nonrelativistic regime, we use eq. (\ref{tlength5}) in eq. (\ref{N2}), resulting
\begin{equation}
	n=\frac{|Q|}{V}
	\label{N3}
\end{equation}
which, up to an exponential small correction that has been omitted, does not depend on $T$. Moreover, it follows also that, when $Q>0$,
$n_{+}$ essentially equals $n$ in eq. (\ref{N3}), whereas $n_{-}$ is exponential small. Now, when $Q<0$, this time is $n_{-}$ that essentially equals $n$ in eq. (\ref{N3}) with $n_{+}$ exponential small. These remarks result from 
%the identities
[cf. eqs. (\ref{charge}) and (\ref{totaln})]
\begin{equation}
\label{identities1}	
N_{+}=\frac{N+Q}{2},
\hspace{1.0cm}
N_{-}=\frac{N-Q}{2},
\end{equation}
and eq. (\ref{densities}). Thus, in the norelativistic regime, one has roughly
either a gas of particles ($Q>0$) or a gas of antiparticles ($Q<0$), with relative densities $n_{+}/n=1$ and  $n_{-}/n=1$, respectively.
Although we should not expect that the present classical toy model is realistic when $T$ is close to absolute zero, it is worth pointing out that as $T\rightarrow 0$ 
eqs. (\ref{cpotential4}) and (\ref{cpotential5}) lead to
$\mu=Mc^{2}$ or $\mu=-Mc^{2}$, corresponding to $Q>0$ and $Q<0$,
respectively. A word should be added regarding eqs. (\ref{boyle}) and (\ref{N})
at the nonrelativistic regime where eq. (\ref{N3}) holds.
The equation of state is essentially given by
\begin{equation}
	PV=|Q|kT.	
	\label{boyle2}	
\end{equation}

By taking into account also  the subleading correction in the 
second expansion in eq. (\ref{abehavior}), eq. (\ref{ienergy2}) yields
the following nonrelativistic internal energy and heat capacity at constant volume:
 \begin{equation}
 	\label{ienergy3}
 	U=|Q|Mc^{2}+\frac{D-1}{2}|Q|kT,\hspace{1.0cm}
 	C_{V}=\frac{D-1}{2}|Q|k.
 \end{equation}
To complete, noticing eqs. (\ref{cpotential4}) or (\ref{cpotential5}), eq. (\ref{N3}) and the text just after it, 
we use eqs. (\ref{boyle2}) and (\ref{ienergy3}) in eq. (\ref{entropy3}) to obtain
\begin{equation}
	\label{st}
	S=k|Q|\ln\left(\frac{V}{|Q|\lambda^{D-1}}\right)
	+\frac{D+1}{2}k|Q|,
\end{equation}
which is the Sackur-Tetrode equation of our toy model.

\subsection{Ultrarelativistic regime: $Mc^{2}\ll kT$}
\label{urregime}
Let us turn now to the  ultrarelativistic  regime
\footnote{As part of our classical toy model we assume that
the gas is very diluted, i.e., $|V/Q|$ is taken to be arbitrary large.
}.
%We consider $T$ high enough such that it is much bigger than the %Fermi temperature (for fermions) or bigger than the critical %temperatue (for bosons) if $h$ were identified with Planck's %constant.}.
By considering  again eqs. (\ref{tlength4}), (\ref{lmultipliers}) and  now the first expression in eq. (\ref{abehavior}),
it results in
%\begin{equation}
%	\Lambda^{D-1}=\frac{(hc)^{D-1}}{2^{D-1}\pi^{D/2-1}
		%\Gamma(D/2)(kT)^{D-1}}.
	%\label{tlength6}
%\end{equation}
\begin{equation}
	\Lambda=\frac{1}{2\pi}\left[\frac{\pi^{D/2}}{\Gamma(D/2)}\right]^{\frac{1}{D-1}}
	\frac{hc}{kT},
	\label{tlength6}
\end{equation}
instead of eq. (\ref{tlength5}).
Now,
by inserting eq. (\ref{tlength6}) into
eq. (\ref{cpotential}), we arrive at the
leading contribution
\begin{equation}
	\mu=\frac{Q}{V}\, \frac{(hc)^{D-1}}{2^{D}\pi^{D/2-1}\Gamma(D/2)(kT)^{D-2}},
	\label{cpotential6}
\end{equation}
which goes to zero as $T$ progressively increases. Correspondingly, the densities 
are essentially given by eq. (\ref{neutralgas}), i.e., $n_{+}=n_{-}=n/2$, with
\begin{equation}
	\label{densities2}	
	n=2^{D}\pi^{(D-2)/2}\Gamma\left(D/2\right)
	\left(\frac{kT}{hc}\right)^{D-1},
	\end{equation}
and where eq. (\ref{tlength6}) has been used.
Alternatively one could have arrived at eq. (\ref{densities2}) by considering eq. (\ref{N}). Note that mass and charge appear only in small corrections to 
the ``Planckian'' density in eq. (\ref{densities2}) which
holds exactly
when $M$ and $Q$
vanish. It is worth remarking that $\mu=QkT/N$ up to corrections.

Considering eq. (\ref{ienergy2}) and taking 
the first expression in  
eq. (\ref{abehavior}) into account leaves us with $U=(D-1)NkT$, which
for $D=4$ resembles eq. (\ref{ru}) but, again, with the important difference that now $N$ depends on $T$.
By using further 
eq. (\ref{densities2}), it results as leading contribution
a ``Planckian'' internal energy, namely
\begin{equation}
	\label{ienergy4}
	U=\frac{2^{D}\pi^{(D-2)/2}}{(hc)^{D-1}}	
	(D-1)V\Gamma\left(D/2\right)
	\left(kT\right)^{D}.
\end{equation}
Now we can use eqs. (\ref{boyle}), (\ref{entropy3}),
and 
  (\ref{cpotential6})
to (\ref{ienergy4}), to derive the following ``blackbody'' identities
\begin{equation}
	PV=\frac{U}{D-1},\hspace{1.0cm} 
	S=\frac{D}{D-1}\, \frac{U}{T},\hspace{1.0cm}
	C_{V}=(D-1)S,
%	\hspace{1.0cm},
	\label{blackbody}
\end{equation}
which are satisfied by leading contributions. 
%in the expressions of these thermodynamic quantities.

\section{Discussion}
\label{discussion}

This paper addressed the thermodynamics of an ideal gas with charge $Q$ that is conserved whereas the total number of particles $N$ varies with $T$. It is a classical toy model in $D$-dimensional Minkowski spacetime that assumes some mechanism of pair production. By working with the microcanonical ensemble we have shown that at the nonrelativistic regime $N$ hardly depend on $T$ and equals $|Q|$, up to exponential small corrections. The corresponding thermodynamics is essentially that of an ordinary ideal gas where the Sackur-Tetrode equation holds
[cf. eqs. (\ref{boyle2}), (\ref{ienergy3}) and (\ref{st})].
In sharp contrast, at the ultrarelativistic regime,
up to negligible corrections, thermodynamics is that of a ``Planck gas'' [cf. eqs. (\ref{ienergy4}) and (\ref{blackbody})] where entropy $S=kDN\propto T^{D-1}$.

We notice that eqs. (\ref{boyle2}) and  (\ref{ienergy3})
are not affected by taking $h\rightarrow 0$ 
(i.e., by considering the small volume $h^{D-1}$ in the phase space
arbitrary small),
whereas such a limit
causes divergences in eqs. (\ref{ienergy4}) and (\ref{blackbody}).
Clearly, this is related to the intrinsic quantum nature of an
ultrarelativistic gas. Namely, if $\Lambda$ in eq. 
(\ref{tlength6}) is considered as a genuine thermal wavelength, by 
noting eq. (\ref{densities2}) one sees quickly that 
$\Lambda$ is comparable to the intermolecular distance $n^{1/(1-D)}$.

It should be remarked that to derive the 
``Planckian'' internal energy in 
eq. (\ref{ienergy4}) we did not take into account if we were dealing with bosons or fermions, since our toy model is classical. Nevertheless, neglecting this fact results that
if we redo the calculations for, say, a Bose gas we will eventually notice a missing factor $\zeta(D)$ in eq. (\ref{ienergy4}), which comes with the use of Bose distributions for particles and antiparticles. In spite of limitations such as this one, it is rather
curious that a classical toy model contains the thermodynamics of
an ordinary ideal gas [see eqs. (\ref{boyle2}), (\ref{ienergy3}) and (\ref{st})]
and that of blackbody radiation
[see eqs. (\ref{ienergy4}) and (\ref{blackbody})] as well.

%%%%%%%%%%%%%%%%%%%%%%%%%Referee's recommendations 1) and 2)
Before closing, we wish to  address further a couple of  issues that are relevant from the physical point of view.
Although our toy model is classical,
%since it uses Boltzmann distribution, 
in a sense it takes into account the Heisenberg uncertainty principle by assuming an arbitrary small but non-vanishing 
volume $h^{D-1}$ in the phase space. Quantum statistical mechanics has another important ingredient that is missing in the toy model. Namely, quantum statistics.
Nevertheless, it seems fair to say that the toy model will 
described reasonably well any relativistic gas with some kind of ``pair production'' mechanism, as long as temperatures involved are not that low as to be comparable   with 
the quantum gases characteristic temperatures. 

There is a point we would like to stress.
The term ``charge'' in this paper should be understood in a broad sense. It means the difference of two species particle numbers regardless of the 
phenomenon responsible for keeping ``charge'' conserved as thermodynamic equilibrium is established. Such a ``pair production'' mechanism 
could be in QED or chemistry, and its details are irrelevant to the final outcome which is the stable thermodynamic equilibrium of the whole system.

%%%%%%%%%%%%%%%%%%%%%%%%%%%%%%%%%%%%%%%%%%%%%%%%%%%%%%%%%%%%%%

\vspace{1cm}
\noindent{\bf Acknowledgements} 

C. P. F. 
acknowledges a grant from
``Coordena\c{c}\~{a}o de Aperfei\c{c}oamento de Pessoal de N\'{\i}vel Superior'' (CAPES).
The work of  E. S. M. Jr. has been partially supported by
``Funda\c{c}\~{a}o de Amparo \`{a} Pesquisa do Estado de Minas Gerais'' (FAPEMIG)
and by ``Coordena\c{c}\~{a}o de Aperfei\c{c}oamento de Pessoal de N\'{\i}vel Superior'' (CAPES).

\appendix 
\section{Some properties of $K_{\nu}(z)$}
\label{appendix}

An integral representation of the modified Bessel function of the second kind, $K_{\nu}(z)$, is \cite{arf85,gra07,mat17}:
\begin{equation}
	K_{\nu}(z)= (z/2)^{\nu}\frac{\Gamma(1/2)}{\Gamma(\nu+1/2)}
	\int_{0}^{\infty}\sinh^{2\nu}(\omega)e^{-z\cosh(\omega)} d\omega
	\label{Macdonald}
\end{equation}
with $Re(\nu)>-1/2$ and $Re(z)>0$.
$K_{\nu}(z)$ satisfies identities such as
\begin{equation}
	zK_{\nu-1}(z)-zK_{\nu+1}(z)=-2\nu K_{\nu}(z)
	%%	\hspace{1.0cm}
	%K_{\nu-1}(z)+K_{\nu+1}(z)=-2\frac{d}{dz}K_{\nu}(z).
	\nonumber	
	%\label{p1}
\end{equation}
and
\begin{equation}
	zK_{\nu+1}(z)=\nu K_{\nu}(z)-z\frac{d}{dz}K_{\nu}(z),
	\hspace{1.0cm}
	zK_{\nu-1}(z)=-\nu K_{\nu}(z)-z\frac{d}{dz}K_{\nu}(z).		
	%\nonumber	
	\label{p1}
\end{equation}
By taking the derivative of eq. (\ref{Macdonald}) and using 
the first identity in
eq. (\ref{p1}),
one obtains the integral representation used in the text, i.e.,
\begin{equation}
	K_{\nu+1}(z)= (z/2)^{\nu}\frac{\Gamma(1/2)}{\Gamma(\nu+1/2)}
	\int_{0}^{\infty}\cosh(\omega)\sinh^{2\nu}(\omega)e^{-z\cosh(\omega)} d\omega.
	\label{Macdonald2}
\end{equation}
We also have the following leading behaviors for small and large values of $z$: %respectively:
\begin{equation}
	K_{\nu}(z\rightarrow 0)=2^{\nu-1}\Gamma(\nu)z^{-\nu}+\cdots,
	\hspace{1.0cm}
	K_{\nu}(z\rightarrow\infty)=\sqrt{\frac{\pi}{2z}}e^{-z}\left[1+
	\frac{4\nu^{2}-1}{8z}
	+\cdots\right],		
	%\nonumber	
	\label{abehavior}
\end{equation}
with $\nu>0$ in the first expression.

\end{document}